\documentclass[aps,reprint,showkeys,noshowpacs,pre]{revtex4-1}
\usepackage{amsmath,amssymb,graphicx}
\newcommand{\ie}{\emph{i.e.}}
\newcommand{\dd}{\mathrm{d}}
\newcommand{\ee}{\mathrm{e}}
\newcommand{\pmax}{p_{\mathrm{m}}}
\newcommand{\eref}[1]{Eq.~(\ref{#1})}
\newcommand{\vek}[1]{\boldsymbol{#1}}
\begin{document}
\title{Firm competition in a probabilistic framework of consumer choice}
\author{Hao Liao$^1$, Rui Xiao$^1$, Duanbing Chen$^{1, 2}$, Mat\'u\v s Medo$^1$, Yi-Cheng Zhang$^1$}
\affiliation{$^1$Physics Department, University of Fribourg, Chemin du Mus\'{e}e 3, CH-1700 Fribourg, Switzerland\\
$^2$Web Sciences Center,School of Computer Science and Engineering, University of Electronic Science and Technology of China, Chengdu 611731, People's Republic of China}

\begin{abstract}
We develop a probabilistic consumer choice framework based on information asymmetry between consumers and firms. This framework makes it possible to study market competition of several firms by both quality and price of their products. We find Nash market equilibria and other optimal strategies in various situations ranging from competition of two identical firms to firms of different sizes and firms which improve their efficiency.
\end{abstract}

\maketitle

\section{Introduction}
Firm competition, one of the most basic market processes, has been famously discussed by Adam Smith~\cite{London1776}. Two pioneering models by Cournot and Bertrand~\cite[Ch.~27]{WWNC2010} then described firm competition by quantity and price, respectively, and provided the first explanations of market behavior in their respective cases. In the Bertrand model, consumers give absolute preference to the lowest price which consequently drives firm profits to zero. By contrast, the Cournot model assumes that the offered products are homogeneous (indistinguishable), derives the price from the aggregate quantity produced by all firms, and allows non-zero profits to be made. While it may seem obvious that Bertrand competition is more beneficial for the consumers than Cournot competition, this is not always the case~\cite{Symeonidis03}. Firm competition models were later improved by modeling the consumer choice through a utility function which is maximized by each individual consumer and whose maximum then reflects the market's behavior. An example of this approach is provided by the classical Dixit-Stiglitz model of monopolistic competition~\cite{DS77} and an extensive line of work that it has inspired~\cite{Brak04}.

It soon became clear that a certain degree of price dispersion is present in real markets~\cite{Stigler61,Rose84} and thus models building on the assumption of a unique price are insufficient. A market where both informed and uninformed customers are present was shown to lead to ``spatial'' price dispersion where some stores sell at a competitive price and others sell at a higher price~\cite{SS77}. The phenomenon of ``temporal'' price dispersion where each store varies its price over time (and thus prevents the customers from learning and distinguishing ``good'' and ``bad'' shops) has been modeled in~\cite{Varian80}. See~\cite{Baye06} for an exhaustive review of work on price dispersion. However, even these models based on the economics of information~\cite{Stigler61} and the search of consumers for information in a market~\cite{EPJB2009} are not entirely satisfactory because they assume that upon inspection, a consumer is able to exactly determine utility of a given product.

We build on a market model where each product is endowed with intrinsic quality and each consumer with quality assessment ability (in general, both quality and ability are continuously distributed over a certain range)~\cite{Zhang2001,PhysicaA2005}. The demand is generated by consumers, not imposed by firms. This model was shown to produce product differentiation where high-quality products target experienced consumers and low-quality products target the unexperienced (or negligent) ones~\cite{EPJB2008}. While~\cite{EPJB2008} deals with the case of heterogeneous consumers served by a monopolist firm, we now focus on homogeneous consumers served by multiple firms. By assuming that each consumer has a maximal price which they are willing to pay, we generalize this framework to include also product price in the consumer decision process. This allows us to model firm competition by product quality and price. With respect to other works where, typically, two consumer groups and two different product levels distinguished by quality or price are considered~\cite{Chioveanu12}, the current framework makes it possible to explicitly study the impact of consumer ability on the market equilibrium. It contributes to an extensive line of complex systems research which has helped to understand basic features of various systems in economics~\cite{MGbook,RMP2009}, sociology~\cite{Castellano09}, and network science~\cite{Newman10}.

The rest of paper is organized as follows. In Section~\ref{sec:basic}, we introduce the framework and find the optimal strategy of a monopolist firm. In Section~\ref{sec:homogeneous}, we study the basic case of two homogeneous firms competing in the market, characterize the market (Nash) equilibrium upon various strategies adopted by these firms, and show that firms can outperform the Nash equilibrium. In Section~\ref{sec:generalizations}, we study three simple generalizations of the basic case: Competition of several firms, competition of firms of different size, and the effect of unequal firm efficiency on the market. In Section~\ref{sec:conclusions}, we summarize the work and discuss the most important open questions and further research directions.

\section{Basic framework}
\label{sec:basic}
\begin{figure}
\centering
\includegraphics[scale=0.35]{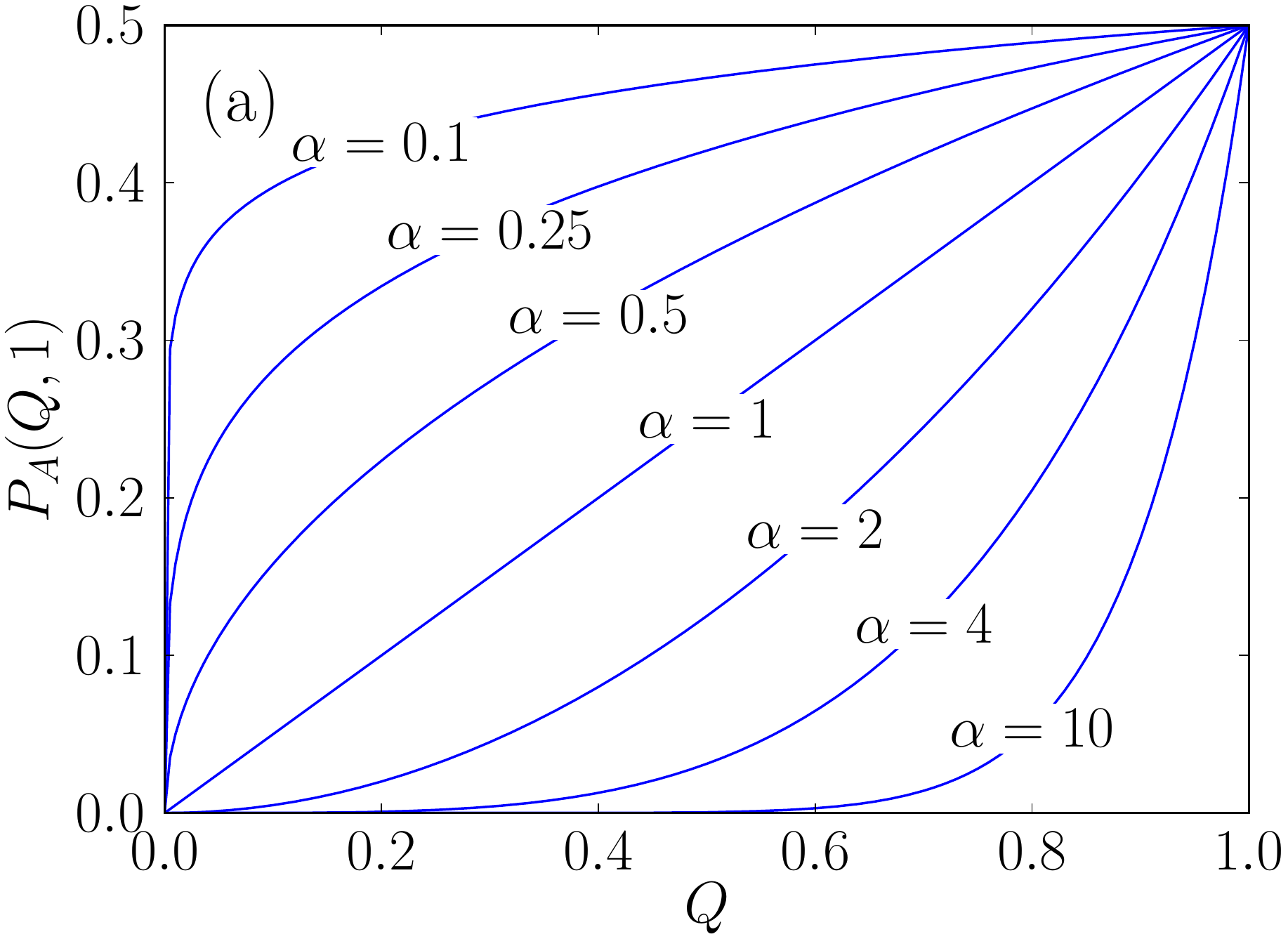}\\[8pt]
\includegraphics[scale=0.35]{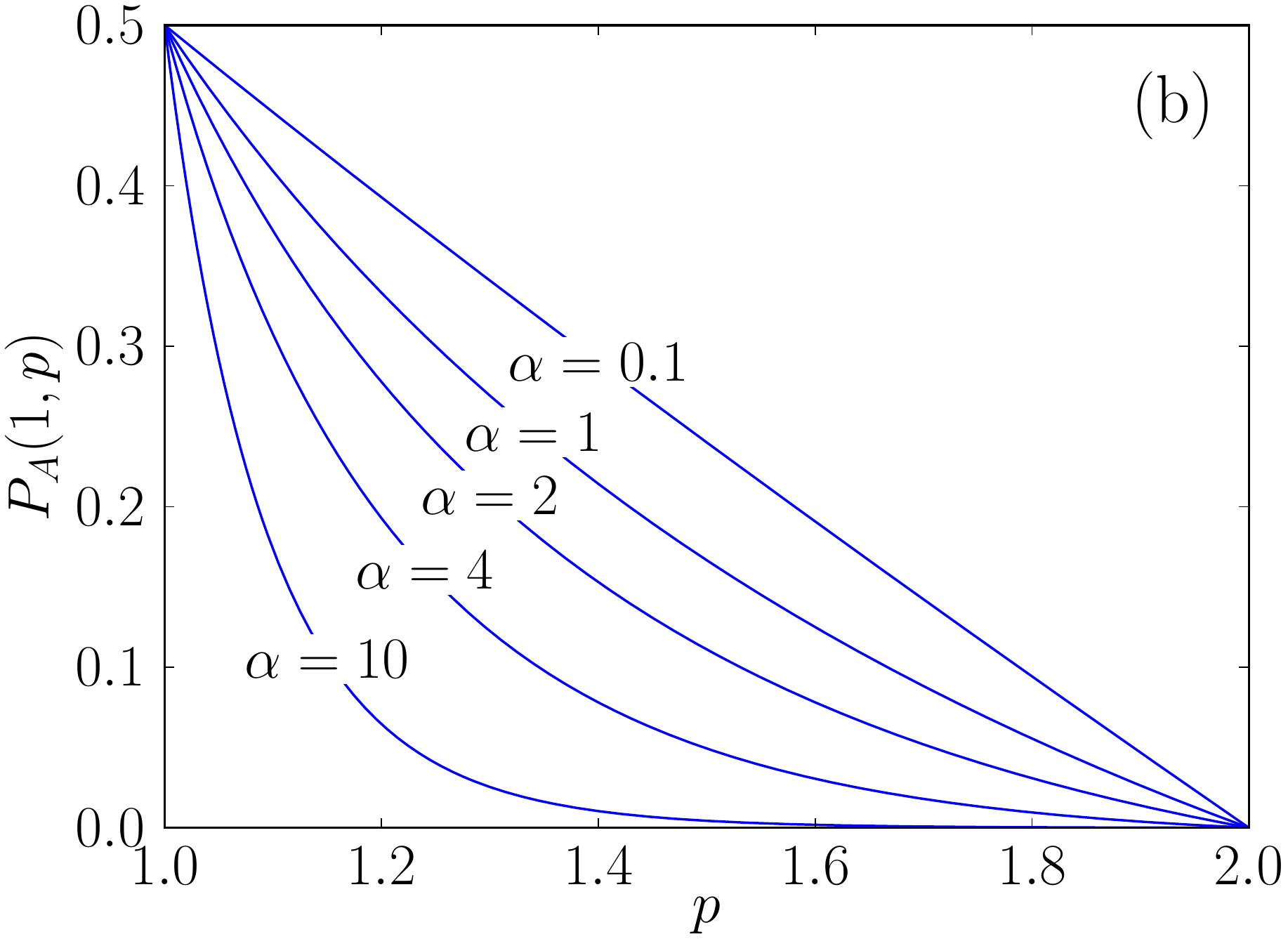}
\caption{Illustrations of the acceptance probability $P_A(Q,p)$ for various values of $\alpha$: (a) Fixed $p=1$ and $\pmax=2$ ($Q$ where profit can be made thus runs from $0$ to $1$), (b) Fixed $Q=1$ and $\pmax=2$ ($p$ where profit can be made thus runs from $1$ to $2$).}
\label{fig:P_A}
\end{figure}

We present here a probabilistic framework for consumer choice and firm profit which deviates from the framework studied in~\cite{EPJB2008} by considering price sensitivity of consumers. The two basic characteristics of any product offered by a firm are its intrinsic quality $Q$ and its price $p$. The probabilistic behavior of the consumers is due to information asymmetry between them and firms which results in consumers being unable to assess quality of products with certainty. The three main assumptions of our framework are as follows:
\begin{enumerate}
\item The firm's profit from each sold unit is assumed in the form $p-Q$. Here $p$ represents the firm's income from selling the item and $-Q$ represents the cost necessary to produce a product of quality $Q$. The use of a different quality-cost relationship does not significantly alter the framework's behavior as long as the basic condition of monotonicity (production cost grows with the product's quality) holds.

\item When offered a product of quality $Q$ and price $p$, the probability that a given consumer accepts the offer and purchases the product---a so-called ``acceptance'' probability---is
\begin{equation}
\label{P_A}
P_A(Q,p)=(1-p/\pmax)(Q/p)^{\alpha}.
\end{equation}
Here the first term reflects that there is a maximal price $\pmax$ that the consumer can afford and the acceptance probability vanishes as $p$ approaches to $\pmax$. The second term reflects the consumer's evaluation of the product quality relative to the product price whereas $\alpha$ is a parameter characterizing how experienced is the consumer. Experienced consumers are able to assess the intrinsic product's quality and their acceptance probability is therefore substantial only when $Q$ is close to $p$ (they require value for their money) which corresponds to $\alpha$ being large. Little experienced consumers are characterized by low $\alpha$ and they are likely to accept also a product with bad $Q/p$ ratio. Plots visualizing the behavior of the acceptance probability given by \eref{P_A} are shown in Figure~\ref{fig:P_A}. Note that Sec.~\ref{appendix1} presents a more fundamental derivation of the acceptance probability. Similarly, the value of $\pmax$ can vary between the consumers and thus reflect their diverse budget constraints. To limit the scope of our present work, we leave the case of heterogeneous $\pmax$ values and their impact on market equilibria for future study.

\item We assume that a consumer facing multiple offers first selects one of them and then decides whether to purchase it or not. It is natural to require that the probability to select a given product---a so-called ``selection'' probability---grows with the product's quality and decrease with the product's price. Since the acceptance probability $P_A(Q,p)=(1-p/\pmax)(Q/p)^{\alpha}$ has exactly these properties, we assume for simplicity that the probability of selecting a particular product is also proportional to $(1-p/\pmax)(Q/p)^{\alpha}$. The separation of consumer decision into a selection step and an acceptance step implies that even when several distinct products are available, at most one of them will be purchased by a given consumer.
\end{enumerate}
While different functional forms can be chosen in each of these three points, our choice represents a simple and yet plausible way to model consumer behavior in a market. Note that we omit a potential $\alpha$-dependent multiplying factor in $P_A(Q, p)$ which was previously used in~\cite{PhysicaA2005,EPJB2008} to reflect the fact that consumers withs low $\alpha$ may reject even a perfect (\ie, $Q=p$) offer. Since we study a homogeneous population of consumers here, this omission has no effect on the found market equilibria. We stress again that the use of consumer acceptance probability implies that demand is generated by consumers, not imposed by firms. On the other hand, quality and price are dictated by firms. This means that out of two classical competition models, that of Cournot and Bertrand~\cite[Ch.~27]{WWNC2010}, our framework is closer to Bertrand competition.

When there is only one firm in the market that offers a product of quality $Q$ and price $p$, rule 3 has no importance and the firm's profit is determined only by the acceptance probability and the profit per sold unit. Firm profit per consumer is therefore $X(Q,p)=(p - Q) P_A(Q, p)$ where $P_A(Q, p)$ needs to be in general averaged over all consumers and their respective parameters $\alpha$ and $\pmax$. Assuming that all consumers are identical, we readily obtain
\begin{equation}
\label{X1}
X(Q,p) = (1-p/\pmax) (Q/p)^{\alpha} (p - Q).
\end{equation}
When $p>\pmax$, this profit is zero because no consumer accepts the offer. When $Q>p$, the firm loses money with each sold item. We thus consider $p<\pmax$ and $Q<p$ as constraints for profit optimization. Note that \eref{X1} is a slight generalization of the profit function in~\cite[Sec.~4]{EPJB2008} where it was implicitly assumed that $\pmax=\alpha+1$ (\ie, experienced consumers are willing to pay high price). The current form with two parameters, $\alpha$ and $\pmax$, makes it possible to model a broader range of consumer behavior than the previous one.

Firms strive to optimize their profit. In the case of one monopolistic firm facing consumers with quality-recognizing ability $\alpha$ and maximal price $\pmax$, profit-maximizing product parameters and the maximal profit per consumer are
\begin{equation}
Q_1^*=\frac{\alpha\pmax}{2(\alpha+1)},\quad p_1^* = \frac{\pmax}2,\quad
X_1^*=\frac{\pmax}{4\alpha}\left(\frac{\alpha}{\alpha+1}\right)^{\alpha}.
\end{equation}
Subscripts denote that these results apply to the case with one firm. One can see that these results are proportional to $\pmax$; this parameter thus not only determines the economically profitable range for $p$ and $Q$ but also the optimal price and quality levels. The same optimal price, quality, and total profit are achieved by several firms acting in the market in collusion. Constrained profit optimization where, for some reason, either quality or price are fixed can be studied as well but we do not report their detailed results here.

\subsection{Two firms or more}
When there are two firms in the market, each offering their own products with quality values $Q_1,Q_2$ and price levels $p_1,p_2$, we use the above-described two-step consumer decision process which was first proposed in~\cite{EPJB2008}. The process consists of two steps: a consumer first selects which product they prefer most (the selection step) and then decides whether to actually buy it or not (the acceptance step). Similarly as the second step is described by the acceptance probability $P_A(Q_i,p_i)$, the first step is described by a so-called selection probability $P_S(Q_i, p_i\vert Q_1,Q_2,p_1,p_2)$ of selecting the product of firm $i$. As explained before, we assume for simplicity that the selection probability is proportional to $(1-p_i/\pmax)(Q_i/p_i)^{\alpha}$. The proportionality constant is given by normalization of this probability (\ie, exactly one product must be selected by each consumer). This approach to the description of the demand of individual consumers facing discrete choices has been used extensively in the past~\cite{McF80,AP92}.

The above-described example is easily generalizable to the case of $N$ firms. Denoting the quality and price values of the offered products as $(Q_1,\dots,Q_N):=\vek{Q}$ and $(p_1,\dots,p_N):=\vek{p}$, respectively, the selection probability becomes
\begin{equation}
P_S(Q_i, p_i\vert\vek{Q}, \vek{p}) = \frac{(1 - p_i/\pmax)(Q_i / p_i)^{\alpha}}
{\sum_{j=1}^N (1 - p_j/\pmax)(Q_j / p_j)^{\alpha}}.
\end{equation}
In summary, the per-consumer profit of firm $i$ which offers a product with quality $Q_i$ and price $p_i$ reads
\begin{equation}
\label{XN}
X(Q_i, p_i\vert \vek{Q}, \vek{p}) = P_S(Q_i, p_i\vert\vek{Q}, \vek{p})P_A(Q_i, p_i)(p_i-Q_i)
\end{equation}
where the selection step, the acceptance step, and the marginal profit from each sold item are combined together. As we shall see in the following section, the introduction of several firms in the market creates a rich system where various game-theoretic concepts apply.

\section{Competition of two homogeneous firms}
\label{sec:homogeneous}
We now proceed to study the behavior of our market model in the presence of two homogeneous firms (\ie, firms sharing all important characteristics such as size and production costs). To maximize their profit, both firms attempt to adjust quality and price of their products in such a way that their own profit is maximized. The situation where both quality and price are the same as in the case of a monopolist firm is therefore inherently unstable: each firm can increase the profit by unilaterally altering quality and price of its product. This would be naturally followed by an analogous move on the competitor's side. The sequence of mutually-provoked adjustments would eventually settle in a configuration where neither side can improve its profit by a unilateral move which is of course the classical Nash equilibrium~\cite{Tadelis13}. The combined profit of the firms is expected to decrease by competition with respect to the monopolist (or non-competitive) case.

\begin{figure}
\centering
\includegraphics[scale=0.35]{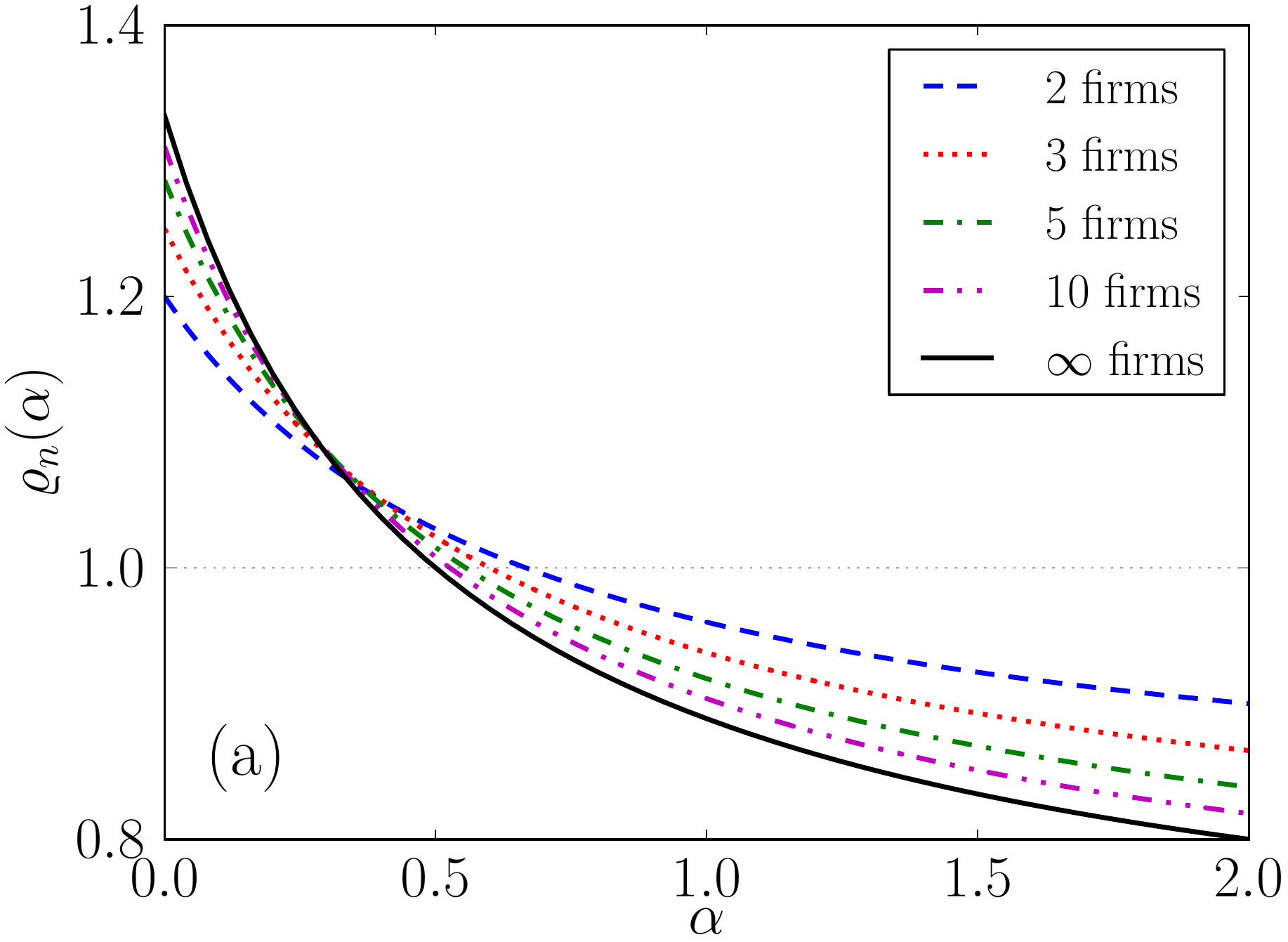}\\[8pt]
\includegraphics[scale=0.35]{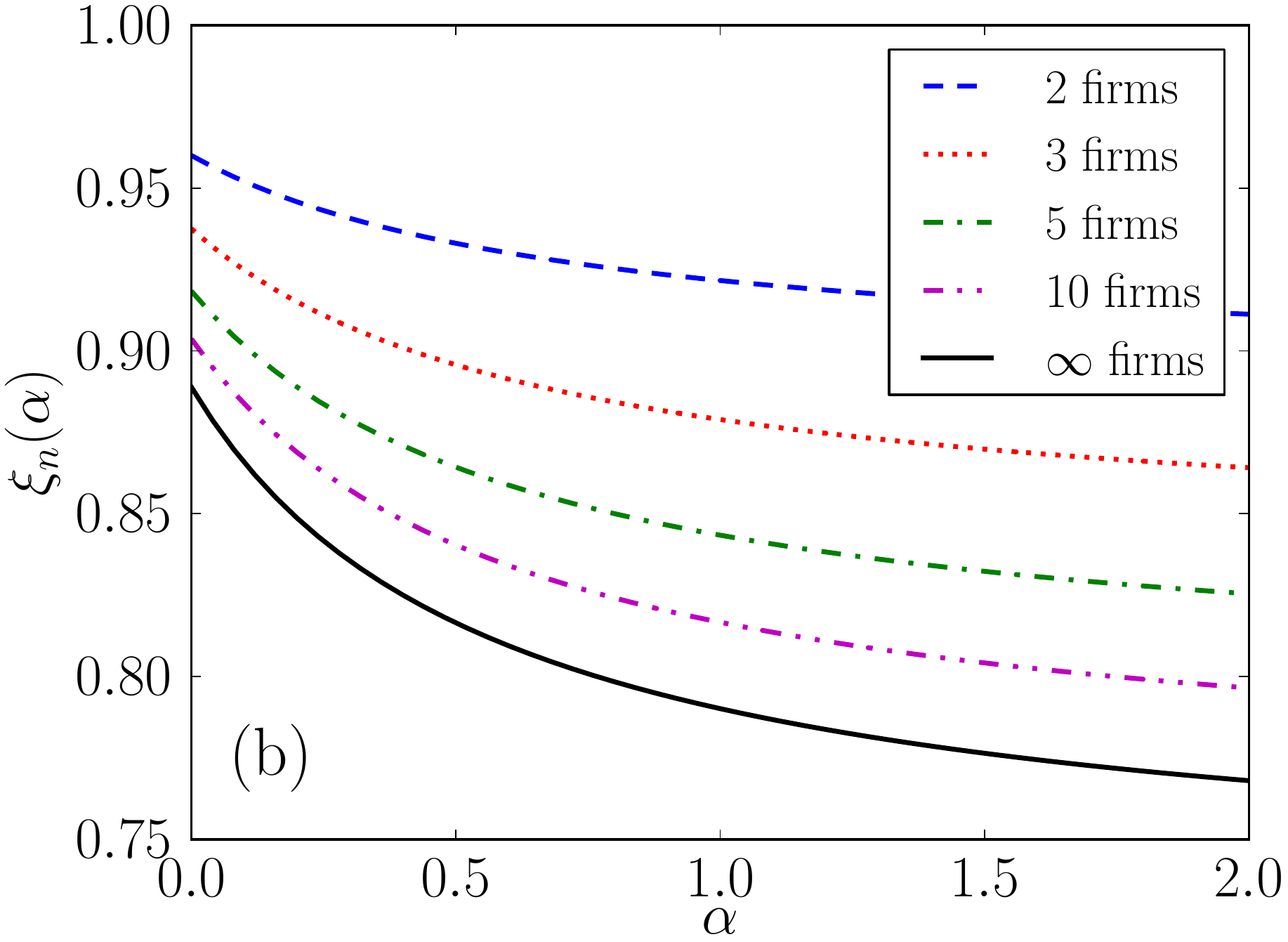}\\[8pt]
\includegraphics[scale=0.35]{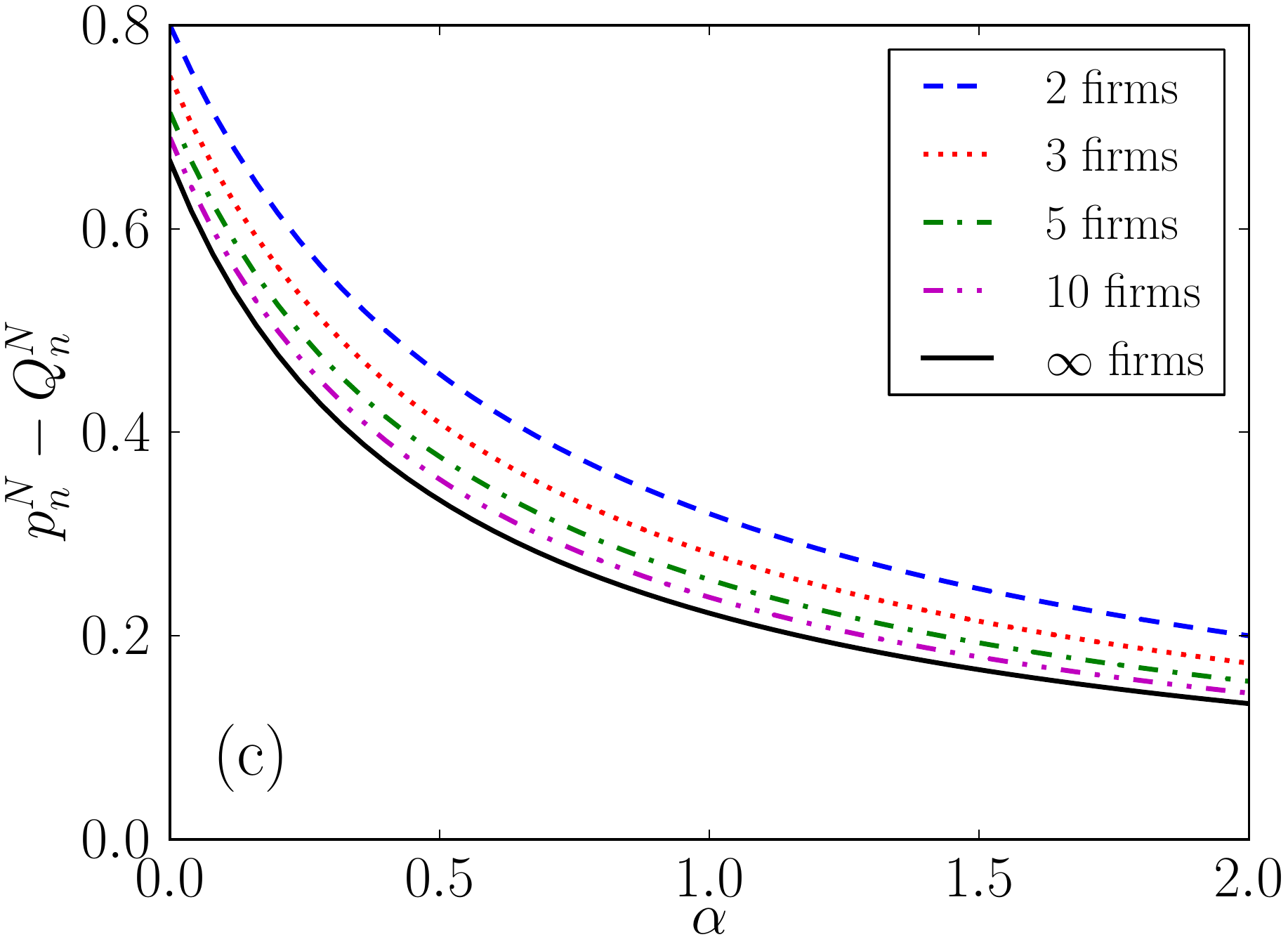}
\caption{A comparison of firm competition outcomes with the one firm case. (a) Quality ratio $\varrho_n(\alpha):=Q_n^N/Q_1^*$ lying both above and below one shows that depending on $\alpha$ and $n$, competition can lead to both increased and decreased quality of products. (b) Profit ratio $\xi_n(\alpha):=X_n^N/(X_1^*/n)$ monotonically decreases with both $\alpha$ and $n$. (c) Marginal profit $p_n^N-Q_n^N$ vanishes when $\alpha\to\infty$ for $n$ fixed but does not vanish when $n\to\infty$ for $\alpha$ fixed.}
\label{fig:n_firms}
\end{figure}

Since there is no reason why quality and price values should differ between the two firms, the Nash equilibrium values $Q_2^N$ and $p_2^N$ must be the same for both firms. These values can be found by requiring that an infinitesimal change by one firm leaves the firm's profit unchanged
$$
X(Q_2^N+\dd Q, p_2^N+\dd p\vert Q_2^N,p_2^N) = X(Q_2^N,p_2^N\vert Q_2^N,p_2^N)
$$
which must hold up the first order in $\dd Q$ and $\dd p$. This is naturally equivalent to $\partial_Q X(Q,p)=\partial_p X(Q,p)=0$ at $(Q_2^N,p_2^N)$. Using \eref{XN}, the market equilibrium can be readily found in the form
\begin{equation}
Q_2^N = \frac{6\alpha\pmax}{5(2+3\alpha)},\quad
p_2^N = \frac25\,\pmax.
\end{equation}
where subscripts denote that the results relate to the case of two competing firms and the superscripts denote that the results represent the Nash equilibrium. It is interesting to note that while product price is \emph{always} lowered by competition, product quality can go either way. The quality ratio
$$
\varrho_2(\alpha) := \frac{Q_2^N}{Q_1^*} = \frac{12(1+\alpha)}{5(2+3\alpha)},
$$
which is plotted in Fig.~\ref{fig:n_firms}a, is greater than one only for $\alpha<\tfrac23$. Similarly, one can compare the resulting profit of each firm with $X_1^*/2$ which is achieved when the two firms do not compete and keep quality and price at the levels corresponding to those of a monopolist firm. The ratio of profit with and without competition, a so-called profit ratio, is for us the key quantity to evaluate the effect of competition on the system. In the given case of two firms competing by both quality and price, it reads
$$
\xi_2(\alpha) := \frac{X_2^N}{X_1^*/2}=\frac{16}{25}\left(\frac{3+3\alpha}{2+3\alpha}\right)^{1+\alpha}.
$$
As shown in Fig.~\ref{fig:n_firms}b, the profit ratio decreases monotonically with $\alpha$ and approaches $16\sqrt[3]{\ee}/25\approx 0.89$ in the limit $\alpha\to\infty$.

When comparing $Q_1^*$ with $Q_2^N$, there is an apparent paradox because quality is improved by competition for unexperienced consumers and worsened for experienced consumers. This is due to the initial optimal setting $Q_1^*=\alpha\pmax/[2(1+\alpha)]$ which is, when $\alpha$ is high, close to the optimal price $p_1^* = \pmax/2$ and thus the marginal profit $p_1^*-Q_1^*$ is small. Price decrease due to competition, if not compensated by a quality decrease, would therefore lead to a negative marginal profit. The optimal firm response to competition is thus to decrease both price and quality. By contrast, when $\alpha$ is small, the difference between $Q_1^*$ and $p_1^*$ is large which makes quality improvement due to competition possible.

\begin{figure}
\centering
\includegraphics[scale=0.35]{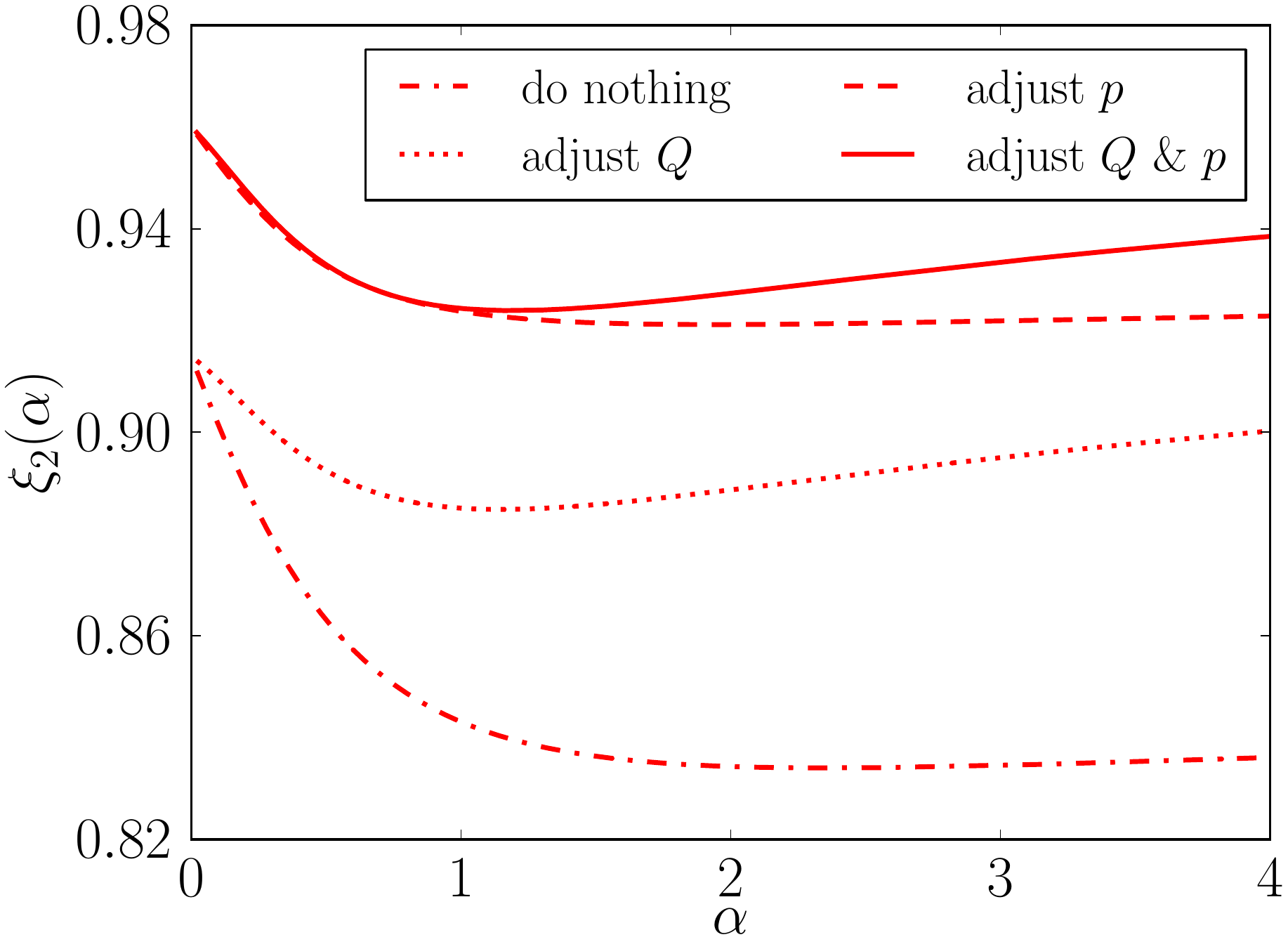}
\caption{The optimal profit ratio $\xi$ of firm 2 as a function of $\alpha$ under four distinct competition strategies.}
\label{fig:firm2}
\end{figure}

It is instructive to study a particular situation where there are two firms in a market and firm $1$ decides to decrease price of its product in order to maximize its profit (product quality is fixed either for production reasons or because of the firm's decision). Firm $2$ then has four different response strategies: Do nothing, adjust price, adjust quality, and finally adjust both price and quality. We focus again on the Nash equilibrium where neither of the firms can improve its profit. The resulting profit ratio of firm $2$ is shown in Fig.~\ref{fig:firm2} as a function of $\alpha$ (the analytical results are too complicated to be shown here). As expected, the first response (do nothing) is the worst one and the last one (adjust both $Q$ and $p$) is the best one. One can note that in the current setting, price competition is more effective than quality competition for all values of $\alpha$. This is a general statement which holds also when it is assumed that firm $1$ has decided to compete by improving product quality or by adjusting both quality and price. The reason lies in the profit formula \eref{X1} which implies that customers are always price-sensitive but their quality sensitivity vanishes as $\alpha\to0$. This is also the reason why for $\alpha\lesssim1$, results achieved by firm 2 with price adjustment are almost as good as those achieved with quality and price adjustment.

\subsection{Beyond the Nash equilibrium}
The Nash equilibrium is a natural competition outcome for two firms focusing solely on maximizing each own's profit given the other firm's strategy. However, the present framework is richer than that. We now consider the case of two firms: while the sole goal of firm 1 is profit optimization, firm 2 is farsighted and does not always try to improve at the expense of its competitor. The motivation for this ``lack'' of competition is that firm 2 knows that even when its profit could be improved by a particular change of strategy, the realization of this change would provoke a strategy adjustment by firm 1 and both firms would eventually end with less profit than before. To model this mechanism, we assume that the farsighted firm 2 chooses the quality and price of its product and keeps it fixed regardless of what does the optimizing firm 1. We consider a simple one-dimensional parametrization
\begin{equation}
Q_2(\tau) = Q_1^* + \tau(Q_2^N - Q_1^*),\quad
p_2(\tau) = p_1^* + \tau(p_2^N - p_1^*)
\end{equation}
which allows firm 2 to continuously adjust between the optimal monopolist behavior (when $\tau = 0$) and the Nash equilibrium of two firms (when $\tau = 1$). The strategy $(Q_1, p_1)$ of firm 1 is chosen as an optimal response to the strategy of firm 2. Profit of both firms is given by \eref{XN}.

\begin{figure}
\centering
\includegraphics[scale=0.35]{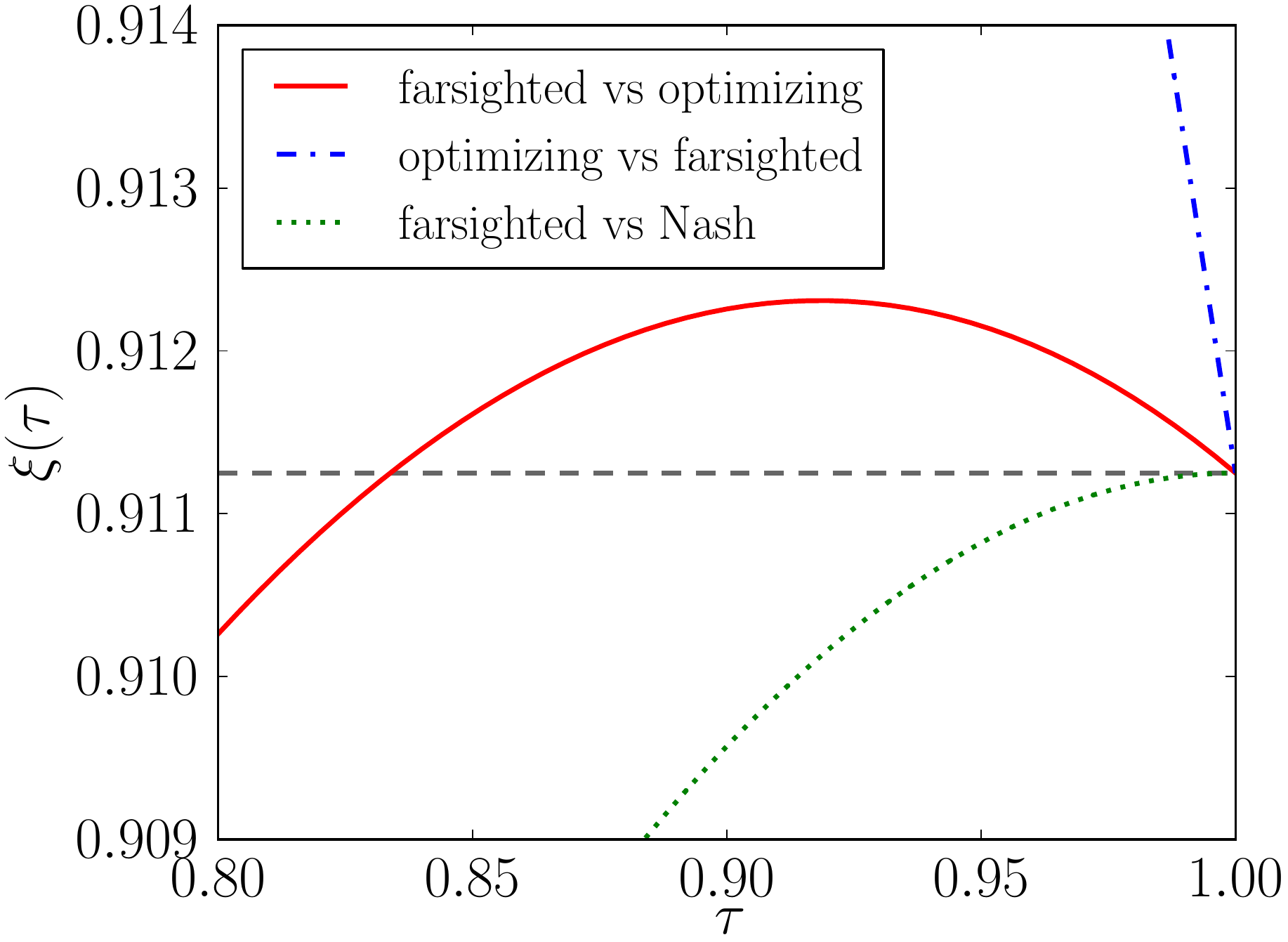}
\caption{The profit ratio as a function of the farsighted firm's strategy for both firms and various competition scenarios; $\alpha = 2$. The horizontal dashed line denotes the Nash profit ratio. While the farsighted firm can improve over the Nash-equilibrium solution over a wide range of $\tau$ (solid line), the optimizing firm benefits more (dash-dotted line).}
\label{fig:beyond}
\end{figure}

As shown in Fig.~\ref{fig:beyond}, labeling firm 2 as farsighted is indeed appropriate because this firm can outperform the Nash equilibrium profit over a wide range of $\tau$ values. The optimal outcome is achieved by $\tau=0.92$ which is significantly different from $\tau=1$ corresponding to the classical Nash equilibrium. The disadvantage of the farsighted approach is that it brings far more benefit to the competing firm 1 which is able to take advantage of the lack of competition on the side of firm 2. Finally, the figure shows also the outcome of the farsighted firm confronted with a firm adopting the Nash equilibrium position. As expected, the farsighted approach yields no extra profit in this case and $\xi(\tau)$ has a unique maximum at $\tau=1$. This demonstrates that the improved profit of the farsighted firm can be realized by lowering the level of competition in the market. When at least one firm competes fiercely by adopting the Nash equilibrium position regardless of the other market participants' actions, no profit improvement is possible. The example of a farsighted firm illustrates that firm strategies other than profit maximization, which are very natural to study in the present framework, should not be neglected.

\section{Generalizations}
\label{sec:generalizations}
Without attempting to be exhaustive, we now discuss three illustrative cases which contribute to the understanding of how customer ability and firm competition affect the market equilibria. Many other modifications and generalizations are possible, some of which are mentioned in Conclusions.

\subsection{Competition of several homogeneous firms}
The treatment presented in Sec.~\ref{sec:homogeneous} can be easily repeated for $n$ competing firms, one of which may attempt to increase its profit by adjusting quality and price of its product. The situation when no profit increase is possible defines the Nash equilibrium with the corresponding quality and price levels
\begin{equation}
\label{n_firms}
Q_n^N = \frac{\alpha n(2n-1)\pmax}{(3n-1)\big(n+\alpha(2n-1)\big)},\quad
p_n^N = \frac{n}{3n-1}\,\pmax.
\end{equation}
When $n\to\infty$, the equilibrium quality approaches $\frac{2\alpha\pmax}{3(2\alpha+1)}$. The ratio of $Q_n^N$ and $Q_1^*$ is
$$
\varrho_n(\alpha) := \frac{Q_n^N}{Q_1^*} = \frac{2n(2n-1)(1+\alpha)}{(3n-1)\big(n+\alpha(2n-1)\big)}
$$
which is greater than one only for $\alpha<\tfrac{n}{2n-1}$ (one can see that as the number of competing firms increases, this threshold decreases to the limit value of $1/2$).

The profit ratio $\xi_n(\alpha) := X_n^N/ (X_1^*/n)$ reads
\begin{equation}
\xi_n(\alpha) = \frac{4n^2}{(3n-1)^2}\left(\frac{(1+\alpha)(2n-1)}{n+\alpha(2n-1)}\right)^{\alpha+1}\!.
\end{equation}
It again decreases with $\alpha$ with the limiting value
$$
\lim_{\alpha\to\infty} \xi_n(\alpha) = 4n^2\exp[(n-1)/(2n-1)]/(3n-1)^2.
$$
As $n$ increases, this itself has a limiting value $4\sqrt{\ee}/9\approx 0.73$ which tells us how much \emph{at most} the total profit decreases by competition (this largest decrease is observed for experienced consumers and many competing firms). We stress that $\xi_n$ is a relative change of profit which already takes the indirect reciprocity with respect to the number of firms into account. The absolute profit of each firm naturally goes to zero as the number of firms increases: Because of competition driving prices down and quality up, but mainly because of splitting the total pie among more participants. To conclude, we show there also the marginal profit $p_n^N-Q_n^N$ which quantifies firm profit per sold item and further shows that while competition does not drive marginal profit to zero, increasing consumer ability does. Lines representing results for $n=3, 5, 10, \infty$ are shown in Fig.~\ref{fig:n_firms} along with the previously found results for two competing firms.

\begin{figure}
\centering
\includegraphics[scale = 0.35]{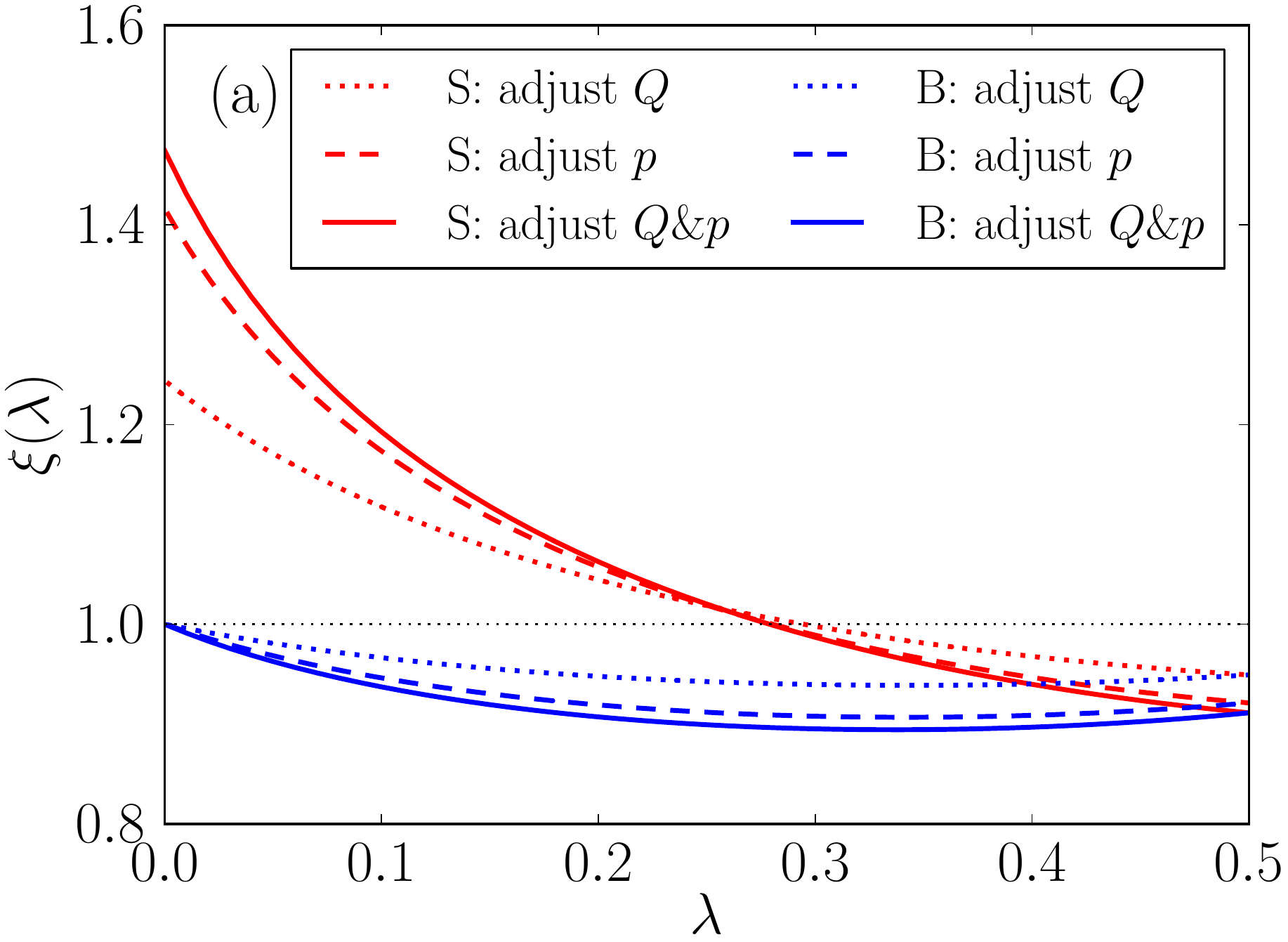}\\[8pt]
\includegraphics[scale = 0.35]{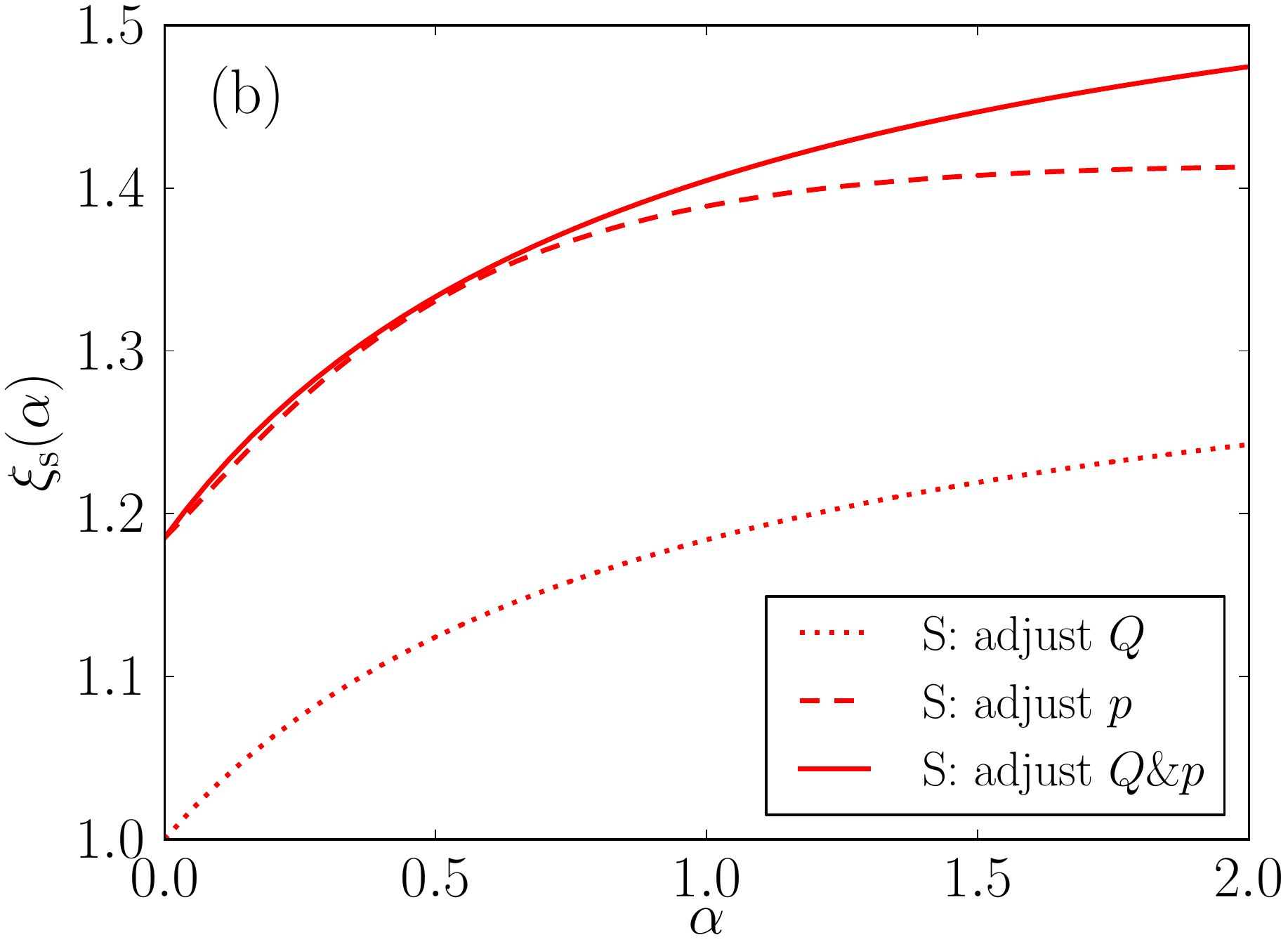}
\caption{Firms of different sizes: (a) The optimal profit ratio vs $\lambda$ for both small (S) and big (B) firm; $\alpha=2$. (b) The optimal profit ratio of the small firm vs $\alpha$ in the limit $\lambda\to0$. In both cases, outcomes of various competition strategies are shown.}
\label{fig:small_firm}
\end{figure}

\subsection{Firms of different sizes}
\eref{XN} implicitly treats both firms as equal---based on quality and price of their products, they both enter in the selection probability and consequently also in the profit equation in exactly the same way. We now generalize this framework by accounting for the case where the visibility of firms differ because of, for example, their different size or different advertising expenditures. Since the two usually go hand in hand (big companies have more resources for advertising), we refer here to a difference in firm size. While there are multiple ways of introducing firm size in the system, we study only the simplest one where firm sizes are exogenous variables which are independent of current firm turnovers and profits (see the concluding section for a further discussion of this issue). If the relative size of firms 1 and 2 are $\lambda$ and $\lambda'=1-\lambda$, respectively, we generalize the probability of a consumer choosing firm 1 over firm 2 to the form
\begin{widetext}
\begin{equation}
\label{PS_small}
P_1(Q_1, p_1\vert Q_2, p_2) = \frac{\lambda (1-p_1/\pmax)(Q_1/p_1)^{\alpha}}
{\lambda (1-p_1/\pmax)(Q_1/p_1)^{\alpha} + \lambda'(1-p_2/\pmax)(Q_2/p_2)^{\alpha}}
\end{equation}
\end{widetext}
The probability of considering the product of firm 2 is complementary, $P_2(Q_2, p_2\vert Q_1, p_1) = 1 - P_1(Q_1, p_1\vert Q_2, p_2)$.

It is straightforward to repeat the previous analysis and study the Nash market equilibrium in this case. As shown in Fig.~\ref{fig:small_firm}a, the small firm can improve its profit by competing with the big one (\emph{i.e.}, by adjusting the quality and price of its product). The limit value of $\lambda$ below which firm 1 can improve its profit by competition varies little with $\alpha$. In the case of competition by both quality and price, for example, this threshold decreases from $0.30$ for $\alpha=0$ to $0.279$ for $\alpha=2$ and $0.275$ for $\alpha=4$. Note that while firm 1 can improve its profit only until certain firm size $\lambda$, the relative decrease of its profit due to competition is always smaller than that of the big firm 2 (due to its greater size, the absolute loss of profit of firm 2 is further magnified with respect to the absolute loss of profit of firm 1). The small firm thus always has the possibility to hurt the big firm by competition more than it does hurt itself. Factors not included in our analysis, such as the volume of firm reserve funds and possible long-term strategic considerations, determine whether firm 1 is ultimately interested in this kind of rivalry or not.

The limit case of a negligibly small firm ($\lambda\to0$) is easy to be treated analytically and leads to the optimal strategy
\begin{equation}
\label{small_firm-result}
p_{\mathrm{s}}^* = \frac{\pmax}3,\quad
Q_{\mathrm{s}}^* = \frac{(\beta-1)\pmax}{3\beta},\quad
\xi_{\mathrm{s}}^* = \frac{16}{27}\left(\frac{\beta+1}{\beta}\right)^{\beta}.
\end{equation}
where $\beta=2\alpha+1$. Since firm 1 is assumed to be negligibly small here, the optimal strategy of firm 2 is that of a monopolist firm. As shown in Fig.~\ref{fig:small_firm}b, the profit ratio of the small firm, $\xi_{\mathrm{s}}^*$, is substantially greater than one for any $\beta\geq1$ ($\alpha\geq0$). Inspecting the small firm's optimal strategy given in \eref{PS_small} more closely, we find that it is based on a considerable price cut and a corresponding quality adjustment which is positive for $\beta<2$ (corresponding to $\alpha<1/2$) and negative for $\beta>2$ (this is the same behavior as we have seen before for $Q_n^N$ which was also greater than $Q_1^*$ only for small $\alpha$). One can similarly study the situation where one small firm enters a market with two firms in the Nash equilibrium---the small firm again achieves higher profit by competing than by ``friendly'' adopting the quality and price level set by its greater rivals. The relative profit improvement of the small firm is then smaller than the values reported in Fig.~\ref{fig:small_firm}. This reduced possible profit is given by $Q_2^N$ and $p_2^N$ leaving less space for the benefit of a third firm than $Q_1^*$ and $p_1^*$ leave for a second firm.

\subsection{The effect of unequal efficiency}
Up to now, we always assumed that the marginal profit of each firm is $p-Q$. We now generalize it to the form $p - \eta Q$ where $\eta$ is a firm-specific parameter which gives us the possibility to model the effect of unequal firm efficiency on the market. In particular, we assume $\eta=1$ to be the default (starting) value which however can be lowered by, for example, optimization of productions processes or by innovations. All other things being equal, smaller $\eta$ implies smaller production cost $\eta Q$ and thus higher marginal profit $p - \eta Q$. One can note that while the upper bound on product price remains to be $\pmax$, the upper bound on product quality becomes $\pmax/\eta$ which grows as $\eta$ decreases. The impact of increased efficiency is therefore two-fold: reduction of the production cost and increase of the maximal economically profitable quality of products.

\begin{figure}
\centering
\includegraphics[scale = 0.35]{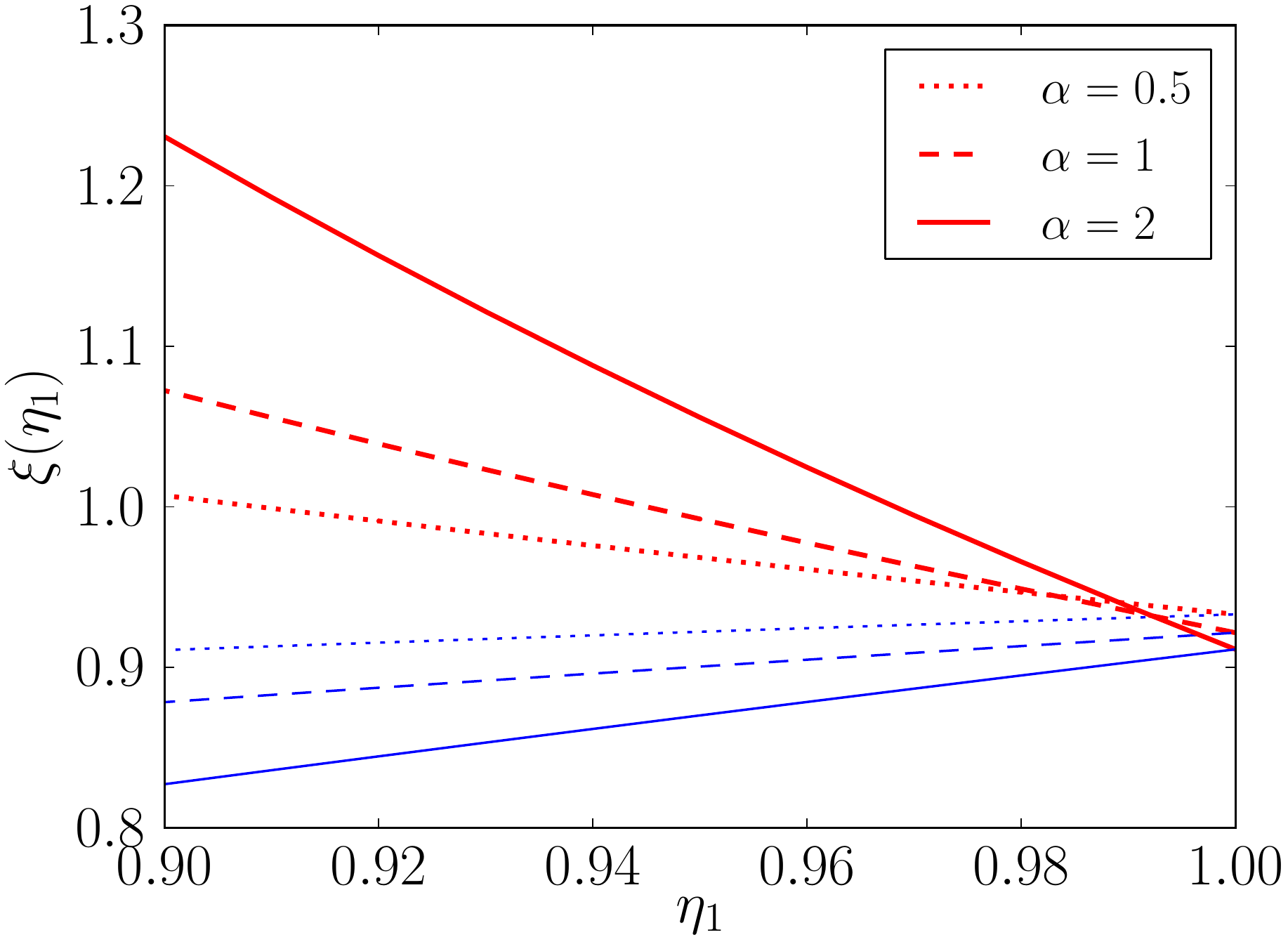}
\caption{The impact of efficiency improvements on firm profit. Firms 1 (improving) and 2 (non-improving) are shown with thick blue and thin red lines, respectively, for three different values of $\alpha$.}
\label{fig:efficiency}
\end{figure}

We study the effect of efficiency on the market on a particular case of two competing firms which are initially homogeneous as they both have $\eta=1$. When firm 1 makes progress and achieves $\eta_1<1$, the market equilibrium ceases to be symmetric (in general, the optimal quality of firm 1 is higher than that of firm 2). As can be seen in Fig.~\ref{fig:efficiency}, the impact on firm profit is particularly strong when $\alpha$ is high. This is because a market with experienced consumers are characteristic by a small marginal profit $p^*-Q^*$. Changing the marginal profit to $p^*-\eta Q^*$ thus results in its substantial relative increase which is in turn followed by an update of the optimal strategy by both firms. The asymmetry between profit changes of the two firms (firm 1 gaining more than firm 2 looses) is not surprising as it reflects an increase of the aggregate profit which is an expected effect of efficiency improvements.

\section{Conclusions}
\label{sec:conclusions}
We have generalized our previously-introduced probabilistic framework for consumer choice based on quality assessment~\cite{EPJB2008} in order to include price sensitivity of consumers and used it to model firm competition in a market. Thanks to this generalization, we have been able to study three different types of competition (by quality, by price, and by both quality and price) and their impact on the market. When several firms compete in a market, it is natural to study the resulting Nash equilibrium where no firm can increase its profit by adjusting properties (quality or price) of its product. Within our framework, product price in the Nash equilibrium is always decreased by competition but product quality can go either way: It increases when consumer ability is low and decreases when consumer ability is high. In addition to the Nash equilibrium, we find that if one firm employs a less aggressive strategy, it can outperform the Nash-optimal profit which is based on the assumption of strict profit maximization by all market participants.

While profit of each individual firm naturally goes to zero as the number of firms grows, their total profit does not vanish due to competition regardless of consumers' ability. Vanishing marginal profits, a classical result of Bertrand competition, are recovered only in the limit of infinitely experienced consumers ($\alpha\to\infty$). This kind of competition is a natural consequence of each firm's attempt to improve its own position which, however, ultimately leads to each firm earning less (similarly as the classical prisoner's dilemma leads to inferior outcome for both prisoners~\cite{Axelrod84}). We find the situation to be different when the competing firms are of a different size. A small firm may even improve its profit by entering in competition with a big firm: The big firm facing a small competitor adjusts the quality and price of its product less than the small firm does which makes a profit increasing-solution possible for the small firm.

The present framework is very elementary, yet it allows for generalizations reflecting various aspects of firm competition in an economy. For example, if multiple firms and multiple groups of consumers with distinct properties are present (as studied in~\cite{EPJB2008}), is it better for each firm to produce one product or is it Nash-stable that each firm produces multiple products---one per each consumer group? There are also research questions that require more substantial modifications of the present framework. The present static framework where firms coexist and their sizes (represented by $\lambda$ and $\lambda'$ in \eref{PS_small}) are fixed. Assuming that the current firm's profit (or turnover) influences the firm's size, the system suddenly gains a temporal dimension and allows one to study effects such as dynamic equilibria, the growth and decay of firms, and the effect of efficiency improvements and innovations in such a dynamical setting. Including the economies of scale has the potential to make the range of produced phenomena even richer. Further insights on which model modifications are indeed crucial and which are not, as well as evidence in support of the basic model, can be gained by attempts to calibrate the present market models on real economic data.

We have simplified the consumer decision process by assuming that each consumer is characterized by one single parameter---quality assessment ability $\alpha$---which then applies to any available product. In reality, however, consumer attention and other resources required for the choice of products are often limited. This limitation is particularly relevant in the nowadays society where there is a multitude of products to choose from~\cite{Dave01}. This can be reflected by endowing each consumer with a limited amount of resources which can be then divided among the available products and spent on their assessment. It is then natural to study questions such as the optimal division of resources and, consequently, the impact of limited consumer attention on the outcome of firm competition in a market.

\bigskip\bigskip
\appendix

\begin{figure}[t]
\centering
\includegraphics[scale = 0.35]{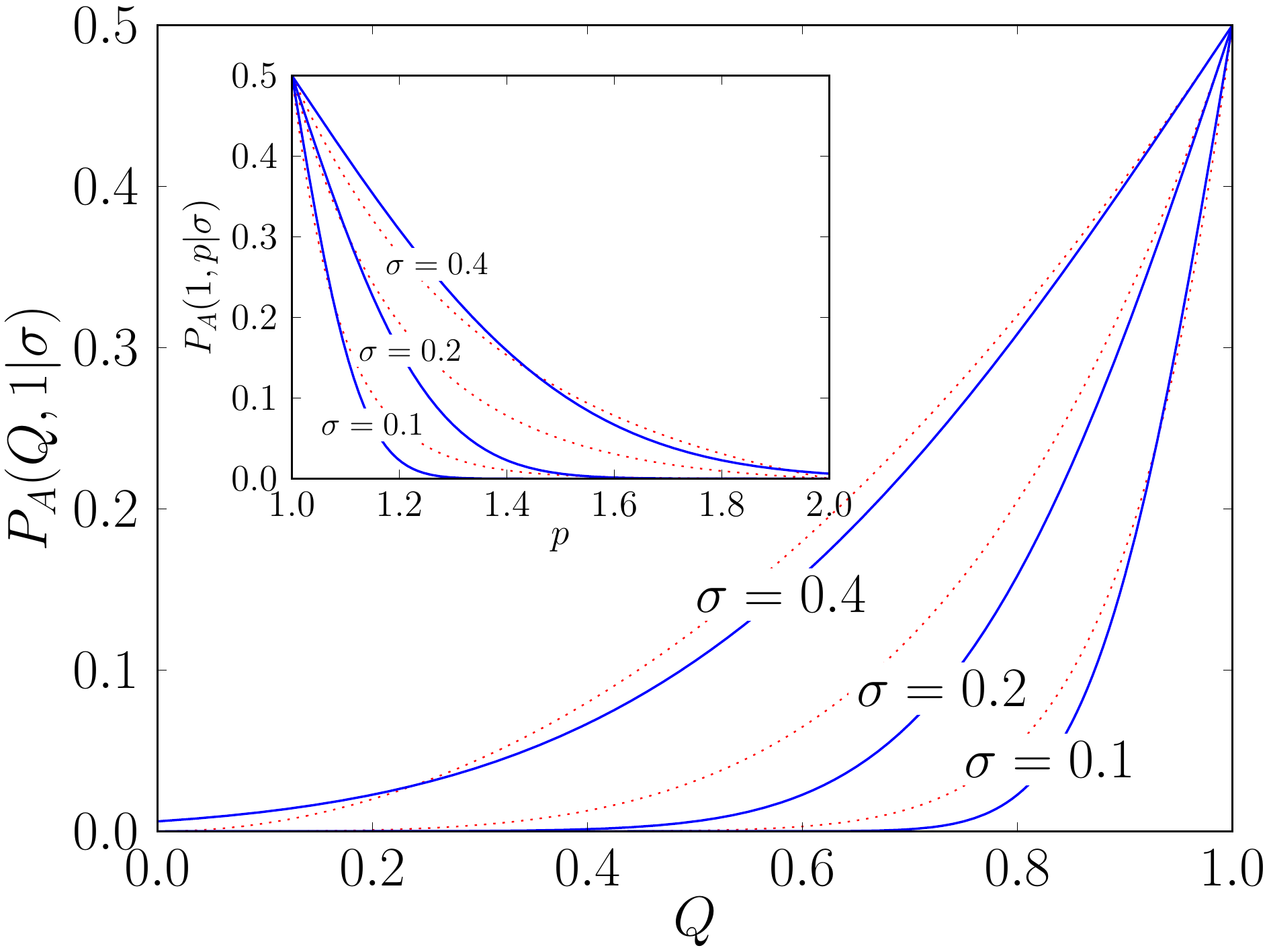}
\caption{The resulting acceptance probability $P_A(Q,p\vert\sigma)$ for various values of $\sigma$ when $p=1$ (main plot) and when $Q=1$ (inset). The thin dotted lines show the acceptance probability course from Fig.~\ref{fig:P_A} for $\alpha=2, 4, 10$.}
\label{fig:appendix}
\end{figure}

\section{On the origin of the acceptance probability}
\label{appendix1}
While it is possible to interpret our consumer decision framework as a purely phenomenological model, we now lay out microscopic foundations for the acceptance probability $P_A(Q,p)$ which is central to the whole framework. We assume that an available product is characterized by its intrinsic quality $Q$ and price $p$. A consumer who evaluates this product is characterized by standard deviation $\sigma$ of perceived product quality $Q'$ which, for the sake of simplicity, is assumed to follow the normal distribution $Q'\in\mathcal{N}(Q,\sigma^2)$. It is a plausible assumption that the consumer accepts the product only if $Q'$ exceeds product price $p$. The acceptance probability can be shown to have the form
$$
P_A(Q,p\vert\sigma) = \frac12-\mathrm{erf}\left(\frac{p-Q}{\sqrt2\sigma}\right).
$$
where $\mathrm{erf}$ is the standard error function. As shown in Fig.~\ref{fig:appendix}, the behavior of this acceptance probability is similar to that of $P_A(Q,p)$ defined by \eref{P_A}. This behavior is recovered over a broad range of underlying assumptions (for example, when additive errors are replaced with multiplicative ones). In this paper, we have chosen to study power-based acceptance probabilities because of their simple form which allows us to obtain analytical results.

\section*{Acknowledgements}
This work was partially supported by the FP7 FET Open research funding scheme through projects Quality Collectives (QLectives, grant no.~231200) and Non-Equilibrium Social Science (NESS, grant no.~296777), and
by the National Natural Science Foundation of China (grant no.~61103109). We thank Matteo Marsili for helpful comments.


\begin{thebibliography}{99}
\bibitem{London1776} A. Smith,
\emph{An Inquiry into the Nature and Causes of the Wealth of Nations} (London, 1776).
\bibitem{WWNC2010} H. R. Varian,
\emph{Intermediate Microeconomics: A Modern Approach} (W. W. Norton \& Company, 8th Edition, 2010).
\bibitem{Symeonidis03} G. Symeonidis,
Comparing Cournot and Bertrand equilibria in a differentiated duopoly with product R\&D,
\emph{Int. J. Indust. Org.} \textbf{21}, 39 (2003).
\bibitem{DS77} A. K. Dixit, J. E. Stiglitz,
Monopolistic Competition and Optimum Product Diversity,
\emph{Am. Econ. Rev.} \textbf{67}, 297 (1977).
\bibitem{Brak04} S. Brakman, B. J. Heijdra (Eds.),
\emph{The monopolistic competition revolution in retrospect} (Cambridge University Press, 2004).
\bibitem{Stigler61} G. J. Stigler,
The Economics of Information,
\emph{J. Polit. Econ.} \textbf{69}, 213 (1961).
\bibitem{Rose84} S. Borenstein, N. L. Rose,
Competition and price dispersion in the US airline industry,
\emph{J. Polit. Econ.} 102, 653 (1995).
\bibitem{SS77} S. Salop, J. E. Stiglitz,
Bargains and ripoffs: A model of monopolistically competitive price dispersion,
\emph{Rev. Econ. Stud.} \textbf{44} 493 (1977).
\bibitem{Varian80} H. R. Varian,
A Model of Sales,
\emph{Am. Econ. Rev.} \textbf{70}, 651 (1980).
\bibitem{Baye06} M. R. Baye, J. Morgan, P. Scholten,
Information, search, and price dispersion,
In: T. Hendershott (Ed.), \emph{Handbook on economics and information systems} (Elsevier, Amsterdam, 2006).
\bibitem{EPJB2009} L. L\"{u}, M. Medo, Y.-C. Zhang,
The role of a matchmaker in buyer-vendor interactions,
\emph{Eur. Phys. J. B} \textbf{71}, 565 (2009).
\bibitem{Zhang2001} Y.-C. Zhang,
Happier world with more information,
\emph{Physica A} \textbf{299}, 104 (2001).
\bibitem{PhysicaA2005} Y.-C. Zhang,
Supply and demand law under limited information,
\emph{Physica A} \textbf{350}, 500 (2005).
\bibitem{EPJB2008} L. L\"{u}, M. Medo, Y.-C. Zhang, D. Challet,
Emergence of product differentiation from consumer heterogeneity and asymmetric information,
\emph{Eur. Phys. J. B} \textbf{64}, 293 (2008).
\bibitem{Chioveanu12} I. Chioveanu,
Price and quality competition,
\emph{Journal of Economics} \textbf{107}, 23 (2012).
\bibitem{MGbook} D. Challet, M. Marsili, Y.-C. Zhang,
\emph{Minority games: interacting agents in financial markets} (Oxford University Press, 2005).
\bibitem{RMP2009} V. M. Yakovenko, J. B. Rosser,
Colloquium: Statistical mechanics of money, wealth, and income,
\emph{Rev. Mod. Phys.} \textbf{81}, 1703 (2009).
\bibitem{Castellano09} C. Castellano, S. Fortunato, V. Loreto,
Statistical physics of social dynamics,
\emph{Rev. Mod. Phys.} \textbf{81}, 591 (2009).
\bibitem{Newman10} M. E. J. Newman,
\emph{Networks: an introduction} (Oxford University Press, 2010).
\bibitem{McF80} D. McFadden,
Econometric Models for Probabilistic Choice Among Products,
\emph{Journal of Business} \textbf{53}, 13 (1980).
\bibitem{AP92} S. P. Anderson, A. de Palma,
The Logit as a Model for Product Differentiation,
\emph{Oxford Economic Papers} \textbf{44}, 51 (1992).
\bibitem{Tadelis13} S. Tadelis,
\emph{Game Theory: An Introduction} (Princeton University Press, 2013).
\bibitem{Axelrod84} R. Axelrod,
\emph{The Evolution of Cooperation} (Basic Books, New York, 1984).
\bibitem{Dave01} T. H. Davenport, J. C. Beck,
\emph{The attention economy: Understanding the new currency of business} (Harvard Business Press Books, 2001).
\end{thebibliography}
\end{document}